\documentclass[aps,prl,twocolumn,superscriptaddress,groupedaddress,footinbib,tightenlines,preprintnumbers]{revtex4}

\usepackage{epsfig}
\usepackage{amstext,amssymb}
\usepackage{amsmath}
\usepackage{graphicx}
\usepackage{dcolumn}
\usepackage[hyperfootnotes=false]{hyperref}
\usepackage{xspace}
\usepackage{braket}
\usepackage{color}
\usepackage{units}
\usepackage{slashed} 

\newcommand{\TeV}{\,\text{TeV}}

\def \P {\mathcal{P}} 
\def \C {\mathcal{C}} 

\definecolor{schrift}{RGB}{100,8,12}

\def \epsilon {\varepsilon} 

      \let\l=\lambda

\let\O=\Omega       
  \let\X=\Xi     
\def\be{\begin{equation}}
	\def\ee{\end{equation}}
\def\bea{\begin{eqnarray}}
	\def\eea{\end{eqnarray}}
\def\bm{\begin{matrix}}
	\def\em{\end{matrix}}
\def\bpm{\begin{pmatrix}}
	\def\epm{\end{pmatrix}}

\begin{document}

\title{\boldmath  \color{schrift}{A minimal model of TeV scale WIMPy leptogenesis }}
\author{Arnab \surname{Dasgupta}}
\email{arnabdasgupta@protonmail.ch}
\affiliation{Institute of Physics, Bhubaneswar, India}
\author{Chandan \surname{Hati}}
\email{chandan@prl.res.in}
\affiliation{Physical Research Laboratory, Navrangpura, Ahmedabad 380 009, India}
\affiliation{Indian Institute of Technology Gandhinagar, Chandkheda, Ahmedabad 382 424, India}
\author{Sudhanwa \surname{Patra}}
\email{sudha.astro@gmail.com}
\affiliation{Center of Excellence in Theoretical and Mathematical Sciences,
Siksha 'O' Anusandhan University, Bhubaneswar-751030, India}
\author{Utpal \surname{Sarkar}}
\email{utpal@prl.res.in}
\affiliation{Physical Research Laboratory, Navrangpura, Ahmedabad 380 009, India}

\begin{abstract}
We present a minimal framework of $U(1)_{B-L}$ gauge extension of the Standard Model explaining dark matter abundance and matter-antimatter asymmetry simultaneously through an attractive mechanism of TeV scale WIMPy leptogenesis, testable at the current and next generation of colliders. This framework can also explain small neutrino masses via a radiative mechanism. One of the key predictions of this model is an enhanced rate for lepton flavor violating decay $\mu \rightarrow e \gamma$ within the sensitivity reach of next generation experiments.
\end{abstract}

\maketitle

{\uppercase\expandafter{\romannumeral 1. \relax}\bf{Introduction---}}
The matter-antimatter asymmetry of the Universe and the nature of non-baryonic dark matter are amongst the most important puzzles in cosmology and particle physics. These are well motivated by the astrophysical observations at a variety of scales and epochs. Observations of the cosmic microwave background (CMB) by WMAP \cite{Hinshaw:2012aka} and PLANK \cite{Ade:2015xua} suggest comparable values of the baryonic and cold dark matter densities
\be
\label{1.1}
\O_{b}h^{2} \sim 0.022, \; \; \; \O_{DM}h^{2} \sim 0.12.
\ee
The standard paradigm in general adopts unrelated mechanisms to explain the observed baryon asymmetry of the Universe (BAU) and the dark matter (DM) abundance, while several co-genesis mechanisms involving asymmetric dark matter (ADM) have also been proposed in literature addressing the comparable abundances of BAU and DM \cite{Davoudiasl:2010am}, often referred to as the ``cosmic coincidence'' problem. In such models a matter-antimatter asymmetry in the dark sector determines the DM abundance and ensures a baryon asymmetry in the visible sector. Thus, ADM scenarios in general require a non-trivial dark sector.

 In this Letter we propose a very minimal framework of $U(1)_{B-L}$ gauge extension of the Standard Model (SM) where the heavy right-handed fields $N_{1,2,3}$, necessary for anomaly cancellation, give the weakly interacting massive particle (WIMP) DM candidates with non-trivial $B-L$ charges ensuring their stability; and their annihilation via lepton number violating out-of-equilibrium scattering processes can realize a very attractive possibility of a TeV scale ``WIMPy leptogenesis", where a lepton asymmetry is first created, which then gets converted into the observed BAU in the presence of sphelaron induced anomalous $B+L$ violating processes before the electroweak phase transition (EWPT). In comparison with the ADM models motivated by the observed $\O_{DM}/\O_{b}\sim 5$ suggesting a common origin for BAU and DM density, the WIMPy leptogenesis is primarily motivated by the WIMP miracle which is extended to explain leptogenesis \cite{Cui:2011ab}. 
 
 A WIMPy model is particularly intriguing because in the WIMP framework a departure from thermal equilibrium occurs naturally and only the violations of lepton (baryon) number, $\C$ and $\C\P$ are required in the DM annihilation processes to satisfy all the Sakharov conditions for leptogenesis (baryogenesis). Another important point worth noting is the presence of washout processes due to inverse annihilations and other lepton number violating scattering processes, which must be suppressed in order to generate a sizable asymmetry. If the washout processes freeze out before the WIMP freeze out then DM annihilations can effectively create a large enough lepton asymmetry which then can give rise to a baryon asymmetry proportional to the WIMP abundance. 
 
Out of the three right-handed fields $N_{1,2,3}$ introduced for anomaly cancellation in $U(1)_{B-L}$ gauge extension of the SM, $N_{1,2}$ with $B-L=-4$ are stable WIMP DM candidates because of the non-trivial $B-L$ charges and after the $U(1)_{B-L}$ is broken by the vacuum expectation value of a singlet scalar field $\zeta$, the DM annihilation via the $L$-violating scattering processes $N_{1,2}N_{1,2} \rightarrow \nu \nu$ mediated through an inert scalar doublet $\eta$ with $B-L=3$ can create a lepton asymmetry. The $\C\P$-violation comes from the interference between tree level scattering process with one-loop diagrams. The inverse annihilation is Boltzmann suppressed for leptogenesis at a temperature $T<m_{DM}$ because the final state leptons in thermal equilibrium are not energetic enough to annihilate back into DM and the washout is dominated by the $L$-violating scattering process $N_{1,2}N_{1,2} \rightarrow Z^{\prime}\rightarrow f\overline{f}$ and $t$-channel $N_{1,2} \bar{\nu}\rightarrow N_{1,2} \nu$. We find that for a $Z^{\prime}$ with mass $\sim$ few $\TeV$, a successful WIMPy leptogenesis can occur at $\TeV$ scale. This is particularly very interesting because unlike the standard leptogenesis case with singlet heavy neutrino or heavy triplet scalar (fermion) decay where the scale of leptogenesis is very high, this model gives an opportunity to verify leptogenesis at the current and next generation colliders. We also discuss a radiative mechanism involving the charged component of the inert scalar doublet $\eta$ which can explain the small neutrino masses confirmed by the atmospheric, solar and reactor neutrino oscillation experiments. We also comment on one of the key prediction of this model being an enhanced rate of the lepton flavor violating (LNV) decay $\mu \rightarrow e\gamma$ which is well within the reach of next generation experiments.\\

{\uppercase\expandafter{\romannumeral 2. \relax}\bf{The Model---}}
In the SM the baryon and lepton numbers are accidental symmetries. The $B$ and $L$ currents are anomalous but the current $B-L$ is an anomaly free combination. The extension of the SM with $U(1)_{B-L}$ induces new gauge as well as gauge-gravitational anomalies in the theory which needs to be cancelled for a self-consistent anomaly free theory. The gauged $U(1)_{B-L}$ gives rise to the triangular gauge anomalies (i) $\mathcal{A}\left[\left(U(1)_{B-L}\right)^3\right]$, (ii) $\mathcal{A}\left[U(1)_{B-L}\left( U(1)_{Y}\right)^2\right]$, (iii) $\mathcal{A}\left[U(1)_{B-L}\left( SU(2)_{L}\right)^2\right]$ and  (iv)$\mathcal{A}\left[\mbox{gravity}^2 \times U(1)_{B-L}\right]$. While the second and the third anomalies are automatically cancelled with the SM field contents, the other two require a +3 contribution coming from additional field content. Conventionally, these anomalies are cancelled by introducing three SM singlet neutral fermion with $B-L=-1$ for each fermion generation; however, such fields are not suitable DM candidates. Another solution to these anomaly cancellations has been pointed out recently with exotic $B-L$ charge assignments to these three neutral fermions in connection with Dirac neutrino masses \cite{Ma:2014qra}. If the three new right handed fields transform as $N_{i}=(-4, -4, +5)$ under $B-L$ symmetry then the first and the fourth anomalies get cancelled: 
\bea
-(-4)^{3}-(-4)^{3}-(+5)^{3} &=& +3\nonumber\\
-(-4)-(-4)-(+5) &=& 3.
\eea
In what follows, we point out an interesting framework using this alternative solution which can give rise to an attractive scenario of WIMPy leptogenesis, where these newly added $N_i$'s can act as WIMPy dark matter. Unlike the generic models of dark matter in the literature, the stability of the dark matter candidate here is ensured automatically because of the non-trivial $B-L$ charges. The particle spectrum of $U(1)_{B-L}$ gauge extension of Standard Model is presented below in Table.\ref{tab:Model}.
\begin{table}[htb!]
\label{tab:Model}
\begin{center}
\begin{tabular}{|c|c|c|c|}
	\hline
			& Field	& $ SU(3)_C \times SU(2)_L\times U(1)_Y$	& $U(1)_{B-L}$	\\
	\hline
	\hline
	Fermions	& $Q_L \equiv(u, d)^T_L$		& $(\textbf{3}, \textbf{2},~ 1/6)$	& $1/3$	\\
			& $u_R$							& $(\textbf{3}, \textbf{1},~ 2/3)$	& $1/3$	\\
			& $d_R$							& $(\textbf{3}, \textbf{1},~-1/3)$	& $1/3$	\\
			& $\ell_L \equiv(\nu,~e)^T_L$	& $(\textbf{1}, \textbf{2},~  -1/2)$	&  $-1$	\\
			& $e_R$						& $(\textbf{1}, \textbf{1},~  -1)$	&  $-1$	\\
			& $N_1$						& $(\textbf{1}, \textbf{1},~   0)$	&  $-4$	\\
			& $N_2$					& $(\textbf{1}, \textbf{1},~   0)$	&   $-4$	\\
			& $N_3$					& $(\textbf{1}, \textbf{1},~   0)$	&   $5$	\\
	\hline
	Scalars	& $H$							& $(\textbf{1}, \textbf{2},~ 1/2)$	&   $0$	\\
			& $\eta$					& $(\textbf{1}, \textbf{2},~   -1/2)$	&   $3$\\
			& $\zeta$					& $(\textbf{1}, \textbf{2},~   -1/2)$	&   $3$	\\  
			& $\chi$					& $(\textbf{1}, \textbf{1},~   0)$	&   $8$	\\ 
			& $\xi$					& $(\textbf{1}, \textbf{1},~   0)$	&   $-1$	\\  
	\hline
	\hline
\end{tabular}
\caption{Particle content for extension of Standard Model with $U(1)_{B-L}$ gauge group 
         and quantum numbers under the relevant gauge groups.}
\end{center}
\end{table}

The relevant interaction Lagrangian of the model is given by
\bea
\label{2.1}
\mathcal{L} &=&y_u\, \overline{q_{L}} \widetilde{H} u_{R} + y_d \overline{q_{L}}  H\, d_{R}
     +y_e\, \overline{\ell_{L}} H e_{R} + y_\nu \overline{\ell_{L}} \eta\, N_{1,2}\nonumber\\
     &+&\sum_{\alpha, \beta=1,2} h_{\alpha \beta} \overline{N^c_{\alpha}} \chi N_{\beta} 
     +\sum_{\alpha, \beta=1,2}  h_{3\alpha} \xi \overline{N^c_{\alpha}} N_3.
\eea
The scalar potential of the model is given by
\bea
\label{2.2}
&&V \left( H, \eta,\zeta, \chi, \xi \right) = \sum_{X=H,\eta,\xi}\left[\mu^2_X \left| X\right|^2 + \lambda_X \left|X\right|^4 \right] \nonumber\\
&+& \sum_{\substack{ X\neq Y\\X,Y=H,\eta,\zeta, \chi, \xi }} \l_{XY}  \left| X\right|^2 \left| Y\right|^2 -\left[\l^{\prime\prime}(\zeta^{\dagger}\eta)^{2}+\rm{h.c.}\right]. \nonumber\\
\eea
Now $B-L$ is broken by the vacuum expectation values (vevs) $\langle \zeta\rangle$ as well as $\langle \chi \rangle$. The former gives rise to the DM annihilation via the $L$-violating scattering processes $N_{1,2}N_{1,2} \rightarrow \nu \nu$ mediated through $\eta$ via the scalar interaction term $\l^{\prime\prime}$ and the latter gives Majorana masses to $N_{i}$'s.

\begin{figure}[h]
\epsfig{file=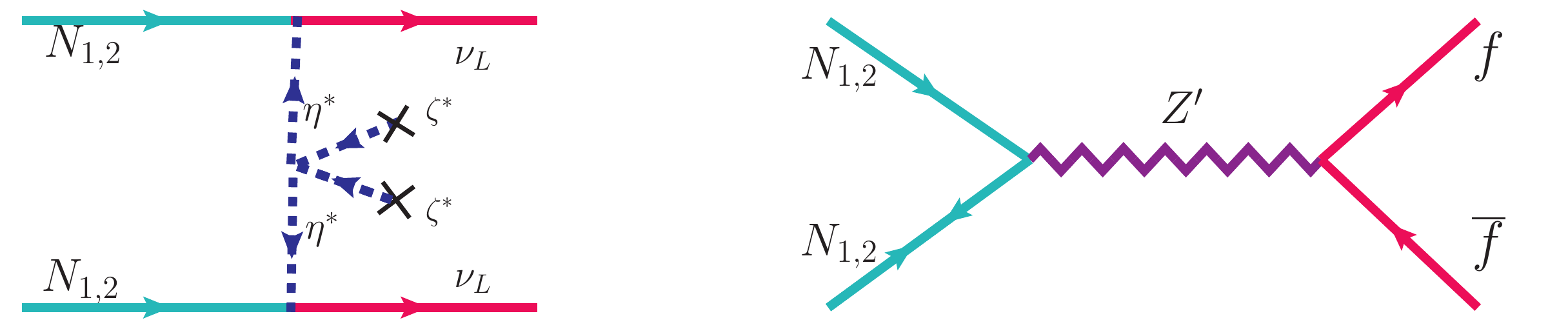,width=0.5\textwidth,clip=}
\caption{Feynman diagrams for WIMPy leptogenesis where left-panel shows the $L$-violating DM annihilation process while 
              right-panel shows the washout  process $N_{1,2}N_{1,2} \rightarrow Z^{\prime}\rightarrow \nu \bar{\nu}$.}
\label{fig1}
\end{figure}

{\uppercase\expandafter{\romannumeral 3. \relax} \bf{WIMPy Leptogenesis---}}
The basic idea of WIMPy leptogenesis relies on the extension of the WIMP miracle framework to accommodate the other ingredients of leptogenesis. The departure from thermal equilibrium is already acquired from the WIMP miracle and one needs to only violate lepton number, $\C$ and $\C\P$ to satisfy all the Sakharov conditions for dynamical generation of a lepton asymmetry. In our model, the stable WIMP DM candidates $N_{1,2}$ with $B-L=-4$ annihilate via $L$-violating scattering processes $N_{1,2}N_{1,2} \rightarrow \nu \nu$ mediated through an inert scalar doublet $\eta$ creating the lepton asymmetry with the $\C\P$-violation coming from the interference between tree level with one-loop diagrams. The relevant diagrams for the tree level DM annihilation process and the washout process $N_{1,2}N_{1,2} \rightarrow Z^{\prime}\rightarrow f\overline{f}$ are shown in Fig. \ref{fig1}. There is another washout process obtained by interchanging the lower legs of the tree level DM annihilation process and using the Majorana-ness of $N_{1,2}$ giving $t$-channel $N_{1,2} \bar{\nu}\rightarrow N_{1,2} \nu$ process which violates lepton number by two units.

The evolution of the lepton asymmetry in WIMPy leptogenesis is manifested in the Boltzmann equation governed by two competing terms: the term corresponding to the annihilation of DM and generation of lepton asymmetry, and the term corresponding to the lepton number violating washout scattering processes driving the lepton asymmetry to equilibrium. Before we write down the formal Boltzmann equations, it is worth noting that the inverse annihilations are Boltzmann suppressed at a temperature $T<m_{DM}$ because the final state leptons in thermal equilibrium are not energetic enough to annihilate back into DM. In the case where $N_{1,2}$ are Dirac fermions the scattering process $N_{1,2}\overline{N_{1,2}} \rightarrow Z^{\prime}\rightarrow f\overline{f}$ do not violate lepton number, however it enters the Boltzmann equation for the number density of $N_{1,2}$. In our model, where $N_{1,2}$ acquires a Majorana mass after $\chi$ acquires a vev breaking $B-L$, the annihilation of $N_{1,2}$ into $f\overline{f}$ violates lepton number and gives a dominant washout process. Thus it is important that this process freezes out before the lepton number violating annihilation process so that a sizable lepton asymmetry can be created. Roughly speaking, for successful generation of lepton asymmetry through the mechanism of WIMPy leptogenesis one should satisfy the following two conditions
\begin{align}
\frac{\langle \sigma_{N_{1,2}N_{1,2} \rightarrow \nu \nu} v \rangle}{\langle \sigma_{N_{1,2}N_{1,2}\rightarrow f\overline{f} }v\rangle}>1,
\end{align}
along with the out-of-equilibrium condition given by
\begin{align}
\frac{n^{N_{1,2}}_{eq}\langle \sigma_{N_{1,2}N_{1,2} \rightarrow \nu \nu} v \rangle}{H}<1,
\end{align}
where $n^{N_{1,2}}_{eq}$ is the equilibrium number density of $N_{1,2}$ and $H$ is the Hubble rate.
\begin{figure}[t!]
\epsfig{file=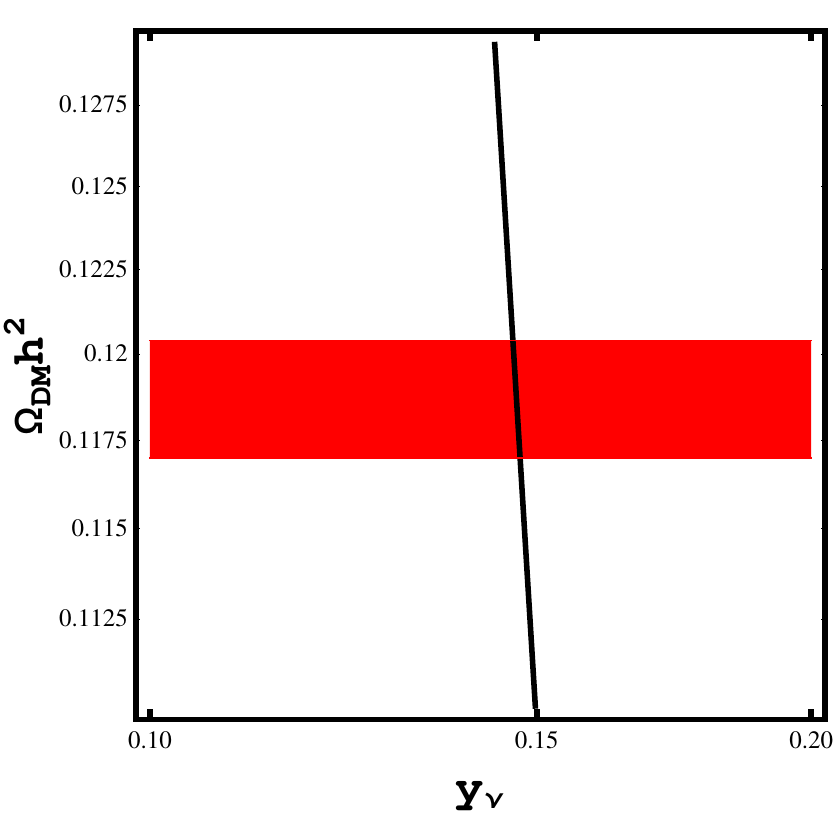,width=0.4\textwidth,clip=}
\caption{Plot showing the variation of DM Relic abundance with the Yukawa coupling. The correct abundance is obtained for the Yukawa coupling $ y_{\nu}\sim 0.145$ for $M_{Z^{\prime}}=5$ TeV and $M_{\eta}=8$ TeV.}
\label{fig3}
\end{figure}
The CP violation is given by
\begin{align}
\epsilon &= \frac{\sigma_{N_{1}N_{1} \rightarrow \nu \nu}-\sigma_{N_{1}N_{1} \rightarrow \overline{\nu} \overline{\nu}}}
{\sigma_{N_{1}N_{1} \rightarrow \nu \nu}+\sigma_{N_{1}N_{1} \rightarrow \overline{\nu} \overline{\nu}}}.
\end{align}
The specific form for the CP asymmetry parameter is given in Appendix A. 

Solving the following Boltzmann equations one can obtain the relic abundance of WIMPy DM after the freeze out of its annihilation and the final lepton asymmetry
\begin{align}{\label{DMBZ}}
& z H(z)s(z)\frac{d Y_{DM}}{dz} = -(\gamma^{CPV} + \gamma^{CPC})\left(\frac{Y^2_{DM}}{(Y^{eq}_{DM})^2}-1\right) \\
& zH(z)s(z)\frac{d Y_{\Delta L}}{d z} = (\epsilon \gamma^{CPV} )\left(\frac{Y^2_{DM}}{(Y^{eq}_{DM})^2}-1\right), \nonumber \\
    &\quad \quad - \gamma^{CPC}\left(\frac{Y^2_{DM}}{(Y^{eq}_{DM})^2}-1\right)
    - \frac{Y_{\Delta L}Y_{DM}}{Y^{eq}_{\Delta L}Y^{eq}_{DM}}\gamma_{WO}, 
    {\label{LBZ}}\end{align}
where $\gamma^{CPV}$ corresponds to CP violating DM annihilation rate, $\gamma^{CPC}$ corresponds to the rate of CP conserving process $N_{1,2}N_{1,2} \rightarrow Z^{\prime}\rightarrow f\overline{f}$, and $\gamma_{WO}$ is the rate of other washout processes such as $N_{1,2} \bar{\nu}\rightarrow N_{1,2} \nu$. The specific forms of the above terms are given in Appendix B. Here $H(z)$ is the Hubble rate, $z$ is the scaled inverse temperature $z\equiv m_{N}/T$ and $Y_{X}\equiv n_{X}(z)/s(z)$, where $s(z)$ is the entropy density.
\begin{figure}[t]
\epsfig{file=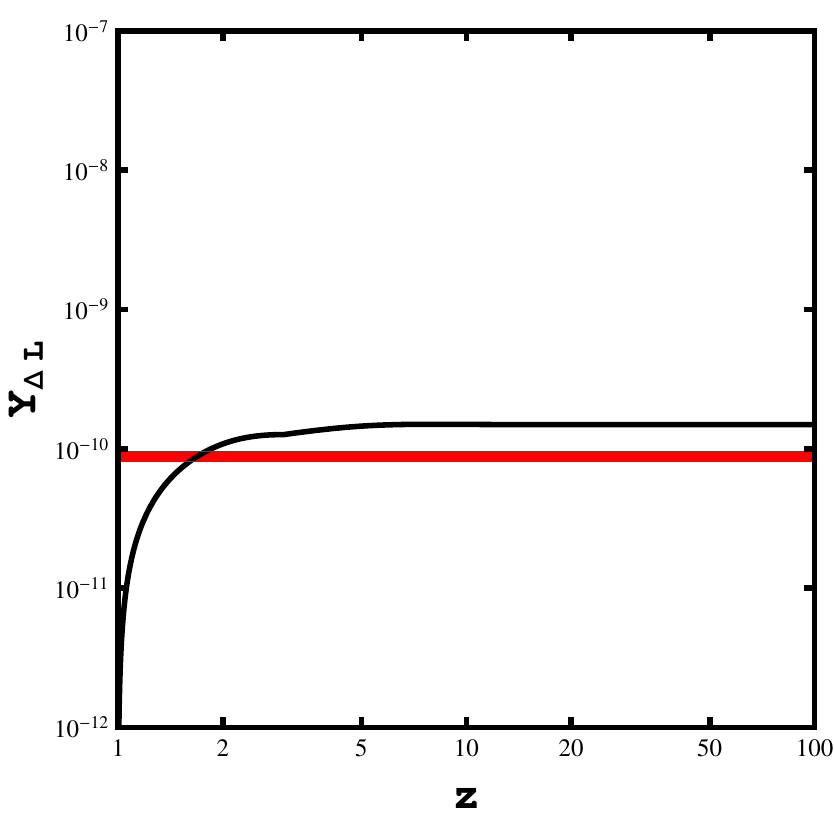,width=0.4\textwidth,clip=}
\caption{Plot showing leptonic asymmetry as a function of inverse temperature ($z=m_{N_1}/T$). The leptonic asymmetry freezes out at around $z\sim 4$ which corresponds to a temerature $T\sim 1.25$ TeV for $m_{N_1}\sim 5$ TeV.}
\label{fig4}
\end{figure}
\begin{figure}[h!]
\epsfig{file=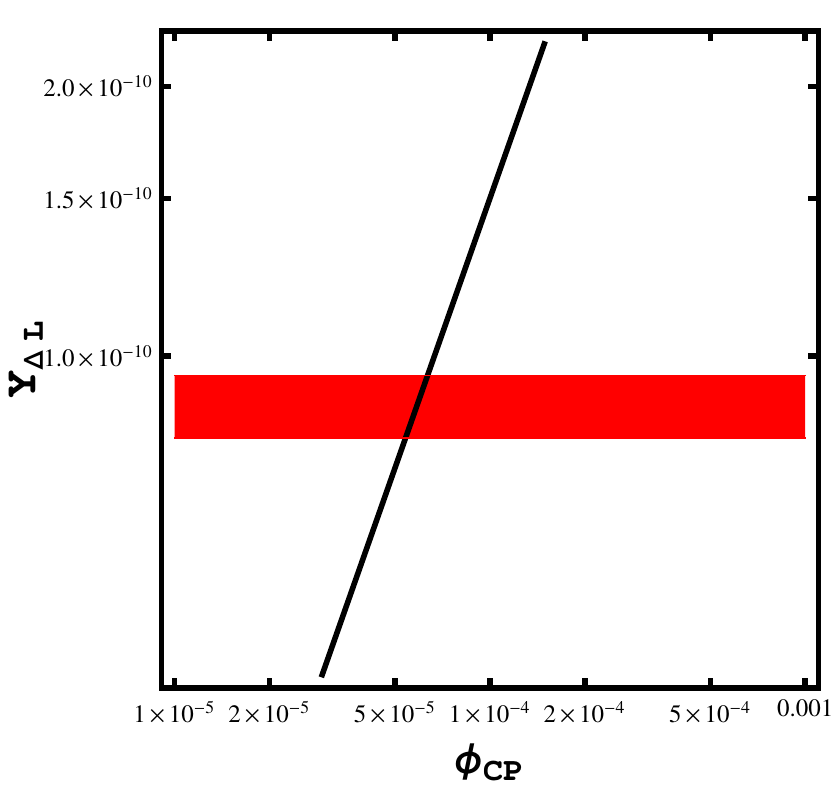,width=0.4\textwidth,clip=}
\caption{Plot showing the variation of the leptonic asymmetry with CP phase $\phi_{\rm CP}$ at $z\sim 100$. The correct asymmetry is generated for the CP phase value $\phi_{CP}\sim 6\times 10^{-5}$.}
\label{fig5}
\end{figure}

Using the Boltzmann equation for DM annihilation given in Eq. (\ref{DMBZ}) one can obtain limits on the Yukawa coefficient $y_{\nu}$ for a given $m_{N}$ (and $M_{Z^{\prime}}$) which can in turn be used as an input to get an approximate solution for $Y_{\Delta L}$. However, it is worth noting that the above is justified under the approximation that the equations for $Y_{DM}$ and $Y_{\Delta L}$ are decoupled up to a first order expansion in asymmetries, i.e., the ``back-reaction" term proportional to $\epsilon Y_{\Delta L}$ contributes negligibly and can be neglected. In Fig. \ref{fig3} we plot the DM abundance as a function of $y_{\nu}$ showing that the correct abundance is obtained corresponding to the Yukawa coupling $y_{\nu}\sim 0.145$ for $M_{Z^{\prime}}=5$ TeV and $M_{\eta}=8$ TeV. The Boltzmann equation governing the lepton asymmetry has two type of terms on r.h.s.; the first term corresponds to generation of lepton asymmetry via DM annihilation proportional to the CP asymmetry and the other two term corresponds to the washout processes depleting the existing lepton asymmetry. We neglect the contribution from inverse annihilation process which is suppressed below the temperature temperature $T<m_{DM}$ and the dominant washout contribution comes from the CP conserving process $N_{1,2}N_{1,2} \rightarrow Z^{\prime}\rightarrow f\overline{f}$ and $N_{1,2} \bar{\nu}\rightarrow N_{1,2} \nu$. In Fig. \ref{fig4} we show the evolution of the lepton asymmetry as a function of $z\equiv m_{N}/T$ showing the freeze out of the lepton asymmetry abundance. Finally, in Fig. \ref{fig5} we plot the final lepton asymmetry as a function of the CP phase $\phi_{\rm CP}$ showing that the correct lepton asymmetry is generated for the CP phase value $\phi_{CP}\sim 6\times 10^{-5}$.

{\uppercase\expandafter{\romannumeral 4. \relax}\bf{Radiative neutrino mass---}}
One of the attractive features of this model is that light neutrino masses can be generated via an one-loop diagram involving $\eta$ and $N_{k}$. This is somewhat similar to the radiative mass mechanism proposed in Ref. \cite{Ma:2006km}, with the key difference being the fact that here the $B-L$ gauge symmetry breaking vev $\langle \zeta \rangle=u$ plays the key role in generating radiative neutrino mass in comparison to the EWSB higgs vev in Ref. \cite{Ma:2006km}. Note that, in Eq. (\ref{2.2}) a Yukawa term involving $\zeta$ similar to the term with coefficient $y_{\nu}$ is assumed forbidden so that at tree level there is no Dirac mass linking $\nu$ and $N$. In addition, the Majorana mass term (after $\chi$ acquires a vev), $\frac{1}{2} m_{k}N_{k}N_{k}+ \rm{h.c.}$, plays a very important role in this mechanism. Assuming that $m_{\eta}\gg m_{k}$, the light neutrino masses at one loop level is given by
\begin{figure}[h!]
\centering
\begin{tabular}{c}
\epsfig{file=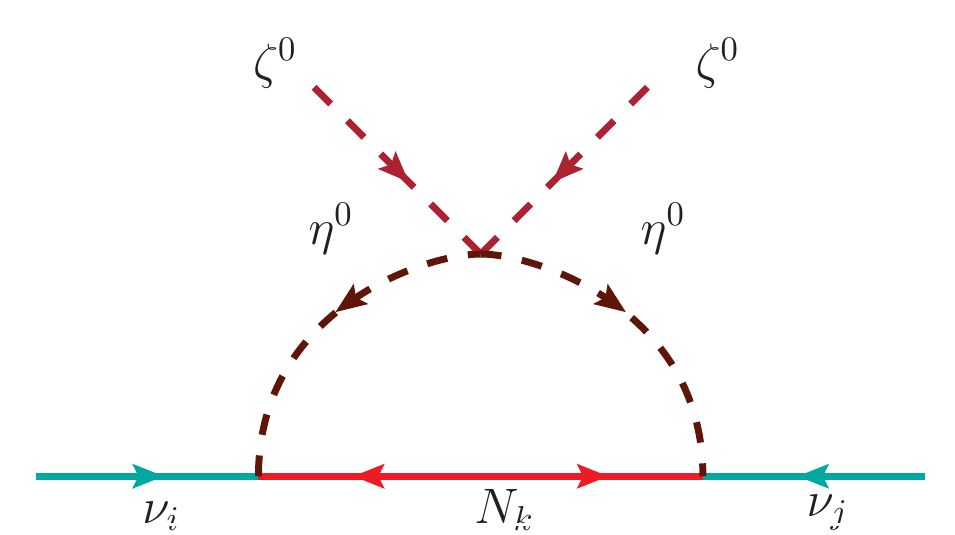,width=0.30\textwidth,clip=}
\end{tabular}
\caption{One loop diagram for radiative light neutrino mass.}
\label{numass}
\end{figure}
\be
\left(m_{\nu}\right)_{ij}\simeq \frac{\l^{\prime\prime}u^{2}}{8\pi^{2}} \sum_{K} {y_{\nu}}_{ij}{y_{\nu}}_{jk}m_{K}.
\ee

\begin{figure}[h]
\centering
\begin{tabular}{c}
\epsfig{file=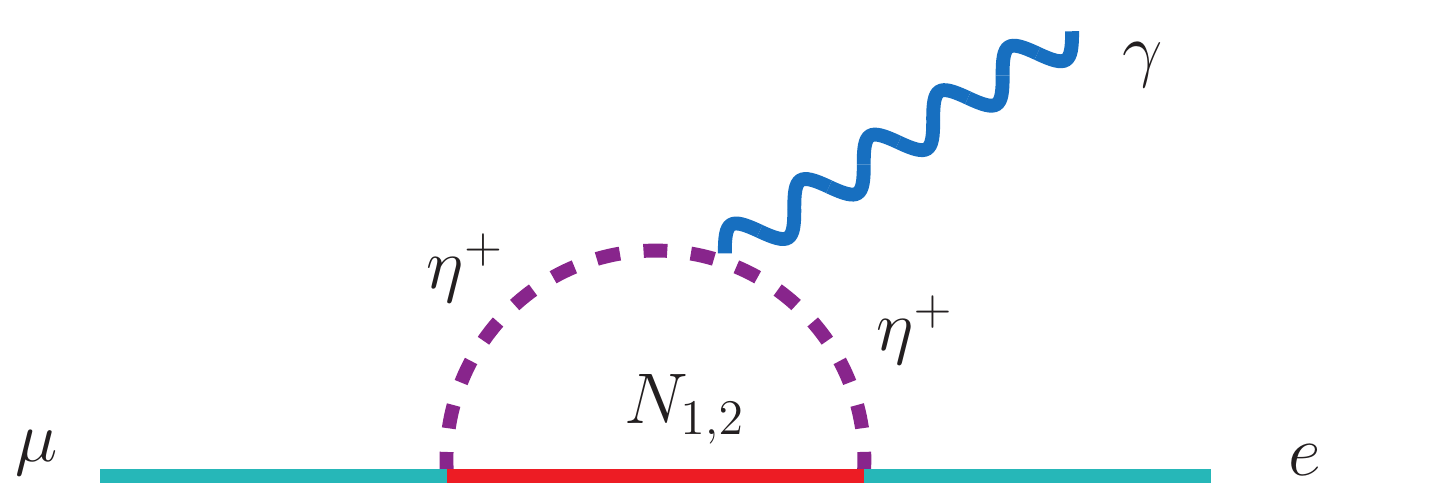,width=0.35\textwidth,clip=}
\end{tabular}
\caption{Diagram showing lepton flavor violating decay $\mu \rightarrow e \gamma$ induced by the charged component of $\eta$.}
\label{meg}
\end{figure}

{\uppercase\expandafter{\romannumeral 5. \relax}\bf{Lepton Flavor Violation---}}
One of the key predictions of this model is an enhanced rate for lepton flavor violating decay $\mu \rightarrow e \gamma$ within the sensitivity reach of proposed next generation experiments. In this model the decay $\mu \rightarrow e \gamma$ can be induced through the diagram shown in Fig.\ref{meg}.
The branching ratio for the lepton flavor violating decay $\mu \to e \gamma$ is given by
$$\mbox{Br}(\mu \to e \gamma) \equiv \frac{\Gamma(\mu \to e \gamma)}{\Gamma(\mu \to e \nu \bar{\nu})} \simeq 5.9 \times 10^{-16} \left(\frac{\mbox{1TeV}}{m_\eta}\right)^4 \left(\frac{y_\nu}{10^{-2}}\right)^4 ,$$
where the individual partial decay widths are given by
\begin{align}
\Gamma_{\mu\rightarrow e \nu \bar{\nu}} &= \frac{G^2_F\, m^5_\mu}{192 \pi^3} I(m^2_e/m^2_\mu),  \nonumber \\
\Gamma_{\mu\rightarrow e \gamma} &= \frac{e^2}{16\pi}\frac{m^5_\mu}{m^4_{\eta}} |y_\nu|^4 |f(m^2_\eta/m^2_{N_{1,2}})|^2, \nonumber \\
\mbox{where} & \nonumber \\
    I(x) &= 1-8\, x -12 x^2 ln(x) + 8 x^3- x^4\, , \nonumber \\
    F(x) &= \frac{1}{16\pi^2}\left[\frac{2x^2-5x-1}{12(x-1)^3} - \frac{x^2\ln x}{2(x-1)}\right] \; .\nonumber 
\end{align}
For $m_\eta \sim$ 10 TeV and $y_\nu \simeq 0.15$, the estimated value for $\mbox{Br}(\mu \to e \gamma)$ is $\sim 3 \times 10^{-15}$. This estimated value of the branching fraction for the decay $\mu \to e \gamma$ is beyond the present experimental sensitivity  $\mbox{Br}(\mu \to e \gamma)\mid_{\rm expt.} \simeq 5.3 \times 10^{-13}$ \cite{Adam:2011ch}; however, it can be probed at the next generation experiments aiming to go down to a sensitivity $\mbox{Br}(\mu \to e \gamma)\mid_{\rm future \; expt} \simeq 10^{-16}$ \cite{Baldini:2013ke}.

{\uppercase\expandafter{\romannumeral 6. \relax}\bf{Conclusions---}}
We have presented a minimal framework of $U(1)_{B-L}$ gauge extension of the Standard Model giving rise to an attractive scenario of TeV scale WIMPy leptogenesis. The heavy right-handed fields $N_{1,2,3}$ necessary for anomaly cancellation can give the correct relic abundance of dark matter and their annihilation via a lepton number violating out-of-equilibrium scattering process mediated by an inert scalar doublet $\eta$ can give rise to an attractive mechanism of a TeV scale WIMPy leptogenesis, testable at the current and next generation of colliders. This framework can also accommodate neutrino mass via radiative mechanism. This model predicts an enhanced rate for lepton flavor violating decay $\mu \rightarrow e \gamma$ within the sensitivity reach of next generation experiments.

\begin{acknowledgements}
{\bf{Acknowledgments---}} AD, CH, and SP would like to thank the organizers of COSMOASTRO15, IOP, Bhubaneswar, India, and WHEPP XIV, IIT Kanpur, India, for their hospitality.  SP would also like to thank Carlos Yaguna for useful discussions and the organizers of the Workshop on Perspectives on the Extragalactic Frontier: from Astrophysics to Fundamental Physics 2016, ICTP, Trieste, Italy.  The work of SP is partially supported by the Department of Science and Technology, Govt.~of India under the financial grant SB/S2/HEP-011/2013. The work US is supported partly by the JC Bose National Fellowship grant under DST, India.
\end{acknowledgements}

\newpage
\appendix*
{\bf{Appendix A---Expression for CP asymmetry}}\\
The tree level cross-section for the $\Delta L \neq 0$ is given as
\begin{align}
|\mathcal{M}^{tree}_c|^2 &=  |y_{\nu}|^4 \lambda''^2v_{\zeta}^4 \sum_{c=u,t}(m_{N_i}^2-c)(m^2_{N_j}-c) \nonumber \\
&\times\left[\frac{1}{(c-
m^2_{\eta_I})^2}-\frac{1}{(c-m^2_{\eta_R})^2}\right]^2,
\end{align}
where $c=u,t$ (Mandelstam variables). The contribution coming from the one-loop diagrams is given by
\begin{align}
\delta^V_{c} &= \sum_{k,\beta}
\frac{4Im[{y_{\nu}}^*_{i\alpha}{y_{\nu}}_{i\beta}{y_{\nu}}^*_{k\beta}{y_{\nu}}_{k\alpha}]}{4 |{y_{\nu}}_{i\alpha}|^2} Im[\Theta(c,m_{\eta_R},m_{\eta_I},m_{N_\beta})] \nonumber \\
& \times \frac{\left[(c-m^2_{\eta_R})+(c+m^2_{\eta_I})\right]}{[(c-m^2_{\eta_R})^2 - (c-m^2_{\eta_I})^2]}(c-m^2_{\eta_R})(c-m^2_{\eta_I}), \nonumber \\
\end{align}
where $m_{\eta R(I)}$ corresponds to the mass of real (imaginary) part of $\eta$ and the loop contribution is given by
\begin{align}
&\Theta(c,m_{\eta_R},m_{\eta_I},m_{N}) = c\left[C(m^2_{N},0,m^2_{\eta_R},0,m^2_{N}) \right. \nonumber \\
&\quad -\left.\left. C(m^2_{N},0,m^2_{\eta_I},0,m^2_{N})\right] \right. \nonumber \\
&\quad - \left. m^2_{N_i}\left[C(m^2_N,m^2_N,0,m^2_{\eta_R},m^2_{\eta_R}) \right.\right. \nonumber \\
&- \left.\left.C(m^2_N,m^2_N,0,m^2_{\eta_I},m^2_{\eta_I})\right] \right. \nonumber \\
&+ \left.c\left[m^2_{\eta_R}D(0,m^2_N,0,m^2_{\eta_R},0,m^2_{\eta_R},m^2_{\eta_R},0,m^2_N) \right.\right. \nonumber \\
&- \left.\left. m^2_{\eta_I}D(0,m^2_N,0,m^2_{\eta_I},0,m^2_{\eta_I},m^2_{\eta_I},0,m^2_N)\right] \right. \nonumber \\ 
&-\left. m^4_{N_i}\left[D(0,m^2_N,0,m^2_{\eta_R},m^2_{\eta_R},0,m^2_N) \right.\right.\nonumber \\ 
&-\left. D(0,m^2_N,0,m^2_{\eta_I},m^2_{\eta_I},0,m^2_N) \right], \nonumber \\
\end{align}
where the Passarino-Veltman functions are defined by
\begin{align}
&D(p^2_1,p^2_2,p^2_3,m^2_1,m^2_2,m^2_3,m^2_4)=\frac{1}{16\pi^4}\int d^4 l \left((l^2-m^2_1)^2\right. \nonumber \\
&\times \left. ((l-p_1)^2-m^2_2)((l-p_2)^2-m^2_3)((l-p_3)^2-m^2_4)\right)^{-1}\nonumber \\
&C(p^2_1,p^2_2,m^2_1,m^2_2,m^2_3)=\frac{1}{16\pi^4}\int d^4 l \left((l^2-m^2_1)^2\right. \nonumber \\
&\times \left. ((l-p_1)^2-m^2_2)((l-p_2)^2-m^2_3)\right)^{-1}\nonumber \\
\end{align}

The asymmetry parameter is defined by
\begin{align}
\epsilon &= \sum_{c=u,t} \delta^V_c|\frac{|\mathcal{M}^{tree}_c|^2}{|\mathcal{M}^{tree}_{total}|^2}
\end{align}
{\bf{Appendix B---Boltzmann Equation for WIMP relic abundance and lepton asymmetry}}\\
The following Boltzmann equations govern the relic abundance of WIMPy DM after freeze out and the abundance of lepton asymmetry.
\begin{align*}
zH(z)s(z)\frac{d Y_{N_i}}{dz} &= - \sum_{j}(\gamma^{\nu\nu}_{N_iN_j} + \gamma^{f\overline{f}}_{N_iN_j})\left(\frac{Y_{N_i}Y_{N_j}}{(Y^{eq}_{N_i}Y^{eq}_{N_j})}-1\right) \nonumber \\
    zH(z)s(z)\frac{d Y_{\Delta L}}{d z} &= \sum_{ij}(\epsilon_{ij} \gamma^{\nu\nu}_{N_iN_j} )\left(\frac{Y_{N_i}Y_{N_j}}{(Y^{eq}_{N_i})Y^{eq}_{N_j}}-1\right) \nonumber \\
    &- \sum_i \frac{Y_{\Delta L}Y_{N_i}}{Y^{eq}_{\Delta L}Y^{eq}_{N_i}}\gamma^{\nu N_i}_{\overline{\nu} N_i} \nonumber \\
    &- \sum_{i,j} \gamma^{f \overline{f}}_{N_{i}N_{j} } \left(\frac{Y_{N_i}Y_{N_j}}{(Y^{eq}_{N_i}Y^{eq}_{N_j})}-1\right)
\end{align*}
where
\begin{align*}
\gamma_{ij}^{kl} &= \int^\infty_{s_inf} \frac{T}{64\pi^2}\sqrt{s}K_1(\sqrt{s}/T)\hat{\sigma}^{ij}_{kl}(s)ds \nonumber \\
\hat{\sigma}^{ij}_{kl}(s) &= \frac{2\lambda(s,m^2_i,m^2_j)}{s}\sigma(s)^{ij}_{kl}   \\
			 &s_{inf} = max\{(m_i + m_j)^2,(m_k + m_l)^2\}. \\
\end{align*}
For different processes $\sigma(s)^{ij}_{kl}$ corresponding to $ij\rightarrow kl$ are given below.
\begin{align*}
\sigma_{N_iN_j}^{f \overline{f}}(s) &= \frac{n_cg^2_fg^2_N}{12\pi s}\left[\frac{1-4m^2_f/s}{1-4m^2_N/s}\right]^{1/2}
\frac{(s+2m^2_f)(s-4m^2_N)}{(s-m^2_Z)^2 + m^2_Z\Gamma^2_Z},  \\
\sigma^{\nu\nu}_{N_iN_j} &= \frac{1}{16\pi \lambda(s,m^2_{N_i},m^2_{N_j})}
\int \sum_{c=u,t}|\mathcal{M}^{tree}_c|^2dt , \\
\sigma^{\nu N_i}_{\bar{\nu} N_j} &= \frac{1}{16\pi \lambda(s,0,m^2_{N_i})}
\int \sum_{c=u,t}|\mathcal{M}^{WO}_c|^2dt ,\\
\end{align*}
where the tree level amplitude squared for DM annihilation is given in Appendix A and the amplitude squared for washout process $N \bar{\nu}\rightarrow N \nu$ is given by
\begin{align*}
|\mathcal{M}^{WO}_c|^2 &= |y_\nu|^4 \lambda''^2 u^4 \sum_{c=u,t}(m_{N_i}^2-c)^2 \nonumber \\
&\times\left[\frac{1}{(c-
m^2_{\eta_I})^2}-\frac{1}{(c-m^2_{\eta_R})^2}\right]^2
\end{align*}
 
\end{document}